\documentclass[10pt,tightenlines,eqsecnum,floats,aps,amsmath,%
amssymb,nofootinbib,prd,showpacs]{revtex4}

\usepackage{amsmath,amsthm,latexsym,amssymb,amsfonts}
\usepackage{enumerate}
\usepackage{graphicx}
\usepackage{calc}
\usepackage{vmargin}
\usepackage{LQC-symbols}

\newcommand{\Z}{{\mathbb Z}}
\newcommand{\R}{{\mathbb R}}

\newcommand{\pp}{{\pi_T}}
\newcommand{\hpp}{\hat{\pi}_T}

\newcommand{\gr}{{\rm gr}}
\newcommand{\scl}{{\rm sc}}
\newcommand{\tot}{{\rm tot}}
\newcommand{\phys}{{\rm phys}}
\newcommand{\Hgr}{{\Hil_{\gr}}}
\newcommand{\Hsc}{{\Hil_{\scl}}}

\newcommand{\Hgrb}{{\Hil_{\gr,B}}}
\newcommand{\Hphys}{{\Hil_{\phys}}}
\newcommand{\Cgr}{{\widehat{C_{\gr}}}}

\newcommand{\Ctot}{{\hat{C}_{\tot}}}
\newcommand{\Span}{{\rm Span}}

\begin{document}

$\hphantom{.}$\vspace{-2cm}
\begin{flushright}
  IGC-08/9-1
\end{flushright}

\title{Physical time and other conceptual issues of QG on the example
  of LQC}

\author{Wojciech Kami\'nski${}^{1}$}
\email{wkaminsk@fuw.edu.pl}
\author{Jerzy Lewandowski${}^{1}$}
\email{lewand@fuw.edu.pl}
\author{Tomasz Paw{\l}owski${}^{2,1,3,4}$}
\email{tomasz@iem.cfmac.csic.es}

\affiliation{
  ${}^{1}$Instytut Fizyki Teoretycznej, Uniwersytet Warszawski,
  ul. Ho\.{z}a 69, 00-681 Warszawa, Poland\\
  ${}^{2}$Instituto de Estructura de la Materia, CSIC,
  Serrano 121, 28006 Madrid, Spain\\
  ${}^{3}$Centrum Fizyki Teoretycznej PAN,
  Al. Lotnikow 32/46, 02-668 Warszawa, Poland\\
  ${}^{4}$Institute for Gravitation and the Cosmos,
  Physics Department, Penn State,
  University Park, PA 16802, U.S.A.
}

\begin{abstract} \noindent{\bf Abstract:\ }
  Several conceptual aspects of quantum gravity are studied on the
  example of the homogeneous isotropic LQC model. In particular:
  $(i)$ The proper time of the co-moving observers is showed
  to be a quantum operator {and} a quantum spacetime metric tensor
  operator is derived. $(ii)$ Solutions of the quantum scalar
  constraint for two different choices of the lapse function are
  compared and contrasted. In particular it is shown that in case
    of model with masless scalar field and cosmological constant
    $\Lambda$ the physical Hilbert spaces constructed for two choices
    of lapse are the same for $\Lambda<0$ while they are significantly
    different for $\Lambda>0$. $(iii)$ The mechanism of the singularity
  avoidance is analyzed via detailed studies of an energy density
  operator, whose essential spectrum was shown to be an interval
    $[0,\rhoc]$, where $\rhoc\approx 0.41\rho_{\Pl}$.
  $(iv)$ The relation between the kinematical and the physical
  quantum geometry is discussed on the level of relation between
  observables.
\end{abstract}

\maketitle

\section{Introduction: the issues raised in the paper}

Loop Quantum Cosmology \cite{lqc1,lqc2} is a family of symmetry
reduced models {built via methods} of Loop Quantum Gravity
\cite{lqg}. It serves {both} as a testing ground for the
quantization frameworks used in Quantum Gravity \cite{lqc2,bahr,kls-obs} and
also a shortcut way to derive some physical predictions. One of
the most surprising predictions it provides is the modification of the
dynamics at near-Planck energy densities leading to the replacement of
the classical Big Bang by a quantum Big Bounce.
Although the most solid and robust results were obtained for
isotropic cosmological models
\cite{aps,aps-imp,apsv-spher,skl-spher,frw-hyper,sLQC,kl-sadj,eff},
there is an ongoing research (with various stages of rigour) treating
homogeneous but anisotropic \cite{B1} or even inhomogeneous models
\cite{eff-als,spher,gm-letter}.
In this paper we are concerned with
some conceptual aspects of quantum gravity and study them on the
example of the homogeneous isotropic LQC model.  They are: existence
of a quantum spacetime metric tensor operator, definition of a
solution to the quantum Einstein constraints, mechanism of
singularity avoidance {and} the role of the kinematical quantum geometry
for the properties of the physical quantum geometry.

Before going to the technical details of the LQC model used in this
work, we {will present an} outline {of} our studies
{(in Sec. \ref{sec:pre-qsp} through \ref{sec:out-zero})}.
Next, in Sec. \ref{sec:LQC} we will introduce the necessary
technical details of the LQC model tested in this work, which is the
model of isotropic, homogeneous spacetime interacting with a
homogeneous scalar field introduced  by Ashtekar, Pawlowski and
Singh \cite{aps-imp}. Most of our results apply also (either
directly or can be generalized) to the so called {solvable} LQC
(sLQC) model \cite{sLQC}.

\subsection{A quantum relativistic time, a quantum spacetime}
\label{sec:pre-qsp}

One of the expectations upon the theory of quantum geometry is that
it should provide a spacetime metric as a quantum operator
\begin{equation}
  \widehat{\rd s^2}\ =\ \widehat{g_{\alpha\beta}}\rd x^\alpha \rd x^\beta\ .
\end{equation}

In the canonical formulation of the Einstein gravity, a general
classical spacetime metric is written in the form
\begin{equation}\label{ds2gen}
  \rd s^2\ =\ -(N^2 - N^a N^b q_{ab}) \rd t^2 
  + N^b q_{ab} (\rd t\rd x^a + \rd x^a\rd t)
  + q_{ab}\rd x^a\rd x^b \ .
\end{equation}
In the gauge choice free approach, the lapse and shift functions $N$
and $N^a$ respectively, are just non-dynamical {gauge
parameters}. Therefore they should pass unchanged to the quantum
theory, allowing in turn to write the metric tensor in the form

\begin{equation}
  \widehat{\rd s^2}\ =\ -(N^2 - N^a N^b \hat{q}_{ab}) \rd t^2
  + N^b \hat{q}_{ab}(t,x) (\rd t\rd x^a + \rd x^a\rd t)
  + \hat{q}_{ab}(t,x)\rd x^a\rd x^b \ ,
\end{equation}
where the un-hatted functions are {independent of} $\hat{q}$. As a
consequence, even in the quantum theory, the $g_{tt}$ metric
 component commutes     with all the other quantum metric
components at any given instant $t$.

However, since Einstein's gravity is a theory with constraints, the
physical Hilbert space differs from the kinematical one, and only
the Dirac observables can give rise to physical quantum observables.
Therefore, the spacetime metric should be first reexpressed in terms
of  {them}. A quite well understood class  {of} the Dirac
observables are the {\it partial observables} developed recently by
Rovelli, Dittrich and Thiemann \cite{rel-obs}. A partial observable
is constructed out of a kinematical observable and a family of
\emph{clock functions} -- functions defined on the classical phase
space providing parametrizations of dynamical trajectories. One of
possible choices of such clock function $T$ is a (coupled to the
gravitational field) Kl{e}in-Gordon massless scalar field (which is
exactly the choice made in the APS model of the quantum FRW
spacetime \cite{aps-imp}). Upon that choice one can write the metric
tensor (\ref{ds2gen}) as,
\begin{equation}
  \rd s^2\ =\ g_{T T} \rd T^2
  + N'^b q'_{ab} (\rd T\rd x^a + \rd x^a\rd T) + q'_{ab}\rd x^a\rd x^b \ .
\end{equation}
The function $g_{TT}$ is of the form
\begin{equation}\label{eq:N'}
   g_{T T} \ =\ -\left( \frac{N^2 - N^a N^b q_{ab}}{N\frac{\pi_T}
   {\sqrt{ {\rm det}q}}+N^aT_a } \right)
\end{equation}
where $\tilde{\pi}_T$ is the momentum canonically conjugate to $T$,
and the second equality follows from the canonical equations
\begin{equation}
  \frac{\partial T}{\partial t}\ =\
  \{\,T,\,\int \frac{N}{2}(\frac{\tilde{\pi}_T^2}{\sqrt{\det q}}
  + q^{ab}\sqrt{{\rm det}q}T_{,a}T_{,b})\ +\ \int N^a\tilde{\pi}_T T_{,a}\} \ .
\end{equation}

From \eqref{eq:N'} it follows immediately that, since all terms on
its righthand side are dynamical quantities, so is  the function
$g_{T T}$. Thus in quantum theory one should consider a Dirac
observable corresponding to it. Whereas on the kinematical Hilbert
space the operators $\hpp$ and $\hat{q}_{ab}$ commute, the
corresponding partial observables do not, therefore the quantum
counterpart of the righthand side of \eqref{eq:N'} is not uniquely
defined. This problem can be seen at the classical level already.
Namely, if we denote by ${\cal O}_{\tilde{\pi}_T}$ and ${\cal
O}_{q_{ab}}$ the corresponding Dirac observables (we suppress the
clock functions and other parameters needed to determine the
observable), then their Poisson bracket \emph{does not vanish}.
Indeed, {(see \cite{GiesThiem-algiv} for details)}
\begin{equation}
  \{ {\cal O}_{\tilde{\pi}_T},\,{\cal O}_{q_{ab}}\}\ =\
     {\cal O}_{\{{\tilde{\pi}_T,q_{ab}}\}^{\rm D}}
\end{equation}
where ${\cal O}_{F(f)} = F({\cal O}_f)$ and $\{\cdot,\cdot\}^D$ is
the Dirac bracket. Furthermore, one can show by inspection {(using
eq. (2.18) of \cite{GiesThiem-algiv})}, that
\begin{equation}
  \{{\tilde{\pi}_T,q_{ab}}\}^{\rm D}\ \not=\ 0 \ .
\end{equation}
In consequence:
\begin{enumerate}[(i)]
  \item a quantum counterpart of $g_{TT}$ is an
    operator which does not commute with $\hat{q}_{ab}$ even at the
    same instant of time,
  \item there is no unique definition of $\hat{g}_{TT}$ because of
    the ordering problem.
\end{enumerate}

In this paper, we point out the issue and propose a definition of the
quantum space-time metric tensor in the APS quantum FRW model, where
the expression for the lapse function \eqref{eq:N'} reduces (due to
homogeneity) to
\begin{equation}\label{N'}
  g_{TT}\ =\ \frac{\sqrt{\det q}}{\tilde{\pi}_T}  \ .
\end{equation}

\subsection{{The physical meaning of the quantum geometry operators}}

{The quantum geometry operators are defined in the kinematical
Hilbert space. They are build, briefly speaking, out of the 3-metric tensor.}
The question regards the role and the properties of the
quantum geometry operators in the physical Hilbert space. Considered
operators can be defined  by using the {relational} observables of
Rovelli-Dittrich-Thiemann. On the one hand, they form in this case the
same Poisson algebra as the kinematical ones. Also in simple examples
($\Lambda=0$) their quantum algebra is equivalent to the algebra of
the kinematical quantum geometry operators. On the other hand, in the
case of $\Lambda>0$ there are many differences between the kinematical
and physical quantum geometry. We discuss them is Sec. \ref{sec:phys-oper}.

\subsection{{Dependence of solutions to the constraint on lapse}}

{In a canonical approach to quantum gravity one has to define
  subsequently
\begin{itemize}
  \item a quantum scalar constraint operator
  \item {the constraint condition, that is the mechanism via
    which the constraint operator selects the physical Hilbert space}
  \item a physical Hilbert space of solutions, {which involves in
    particular specification of the scalar product on it.}
\end{itemize}
In our case} the quantum constraint operator has the form
\begin{equation}\label{eq:Corig}
  \Ctot(N)\
  =\ N(\frac{1}{2}{\hpp^2}\widehat{\sqrt{\det q}^{\ -1}}\ +\ \Cgr)
  \ ,
\end{equation}
where $N$ is the lapse. One choice is to take lapse to be a number.

On
the other hand, taking into account (\ref{N'}) and the quantum nature
of the lapse one is lead to {the constraint} operator
\begin{equation}\label{eq:Ctheta}
  \Ctot(N')\ =\
  \frac{1}{2}{\hpp}\ +\ \hpp^{-1}\left[\widehat{{\sqrt{\det
      q}\,}^{-1}}\right]^{-1}\Cgr \ ,
\end{equation}
{suitably symmetrised in the second term}\footnote{{For the
sake of generality, we distinguish between $\widehat{{\sqrt{\det
q}\,}^{-1}}$ and  $\widehat{{\sqrt{\det q}\,}}^{-1} $. This
distinction takes place if one wants to derive the APS model by the group
averaging method. However our results apply also to the sLQC in which
there is no distinction of this type.}}

Given either one of the constraints,
{one can turn to the second step and define the corresponding
constraint condition. It reads: take  the spectral decomposition
defined by the operator and allow only elements of the Hilbert space
corresponding to the zero eigenvalue.}

{At this point we make a suprising discovery:
\begin{itemize}
  \item On the one hand,  the first operator (\ref{eq:Corig}) has a
    unique self-adjoint extension for arbitrary cosmological constant;
    we point it out in Section \ref{sec:C-noneq} and give a
    mathematical argument.
  \item On the other hand, the second operator (\ref{eq:Ctheta}) has
    inequivalent self-adjoint extensions if
    $\Lambda>0$. \footnote{{This property
    (and its consequences) will be presented in detail in
    \cite{posL1,posL2} currently in preparation.}}
\end{itemize}}

{In other words, the second constraint operator does not
define a constraint condition  uniquely, because the spectral
decomposition depends of a self-adjoint extension. Hence, solutions
to that quantum scalar constraint depend on some additional choice
which has to be made. {This apparent discrepancy forces us to
ask a question:} What is a relation between the uniquely defined
Hilbert space of solutions of the  constraint (\ref{eq:Corig}) and
the extension dependent Hilbert spaces of solutions to the second
constraint (\ref{eq:Ctheta})?}

{To address it} {we explain in detail the difference in
the
  properties of the
operators \eqref{eq:Corig} and, respectively, \eqref{eq:Ctheta}  in
Sec. \ref{sec:C-noneq}.
Also, we briefly discuss the relation between Hilbert spaces of
the solutions to those different constraints
(the detailed analysis will be presented in \cite{ga}).}

\subsection{{Big-Bounce and the energy density operator}}

{Within cosmological model specified at the end of  Section
  \ref{sec:pre-qsp} the
equality satisfied by the lapse function (now given by \eqref{N'}) can
be also written in the following way
\begin{equation}
  N'^2\rd T^2\ =\ 2\rho^{-1} \rd T^2
\end{equation}
where
\begin{equation}
  \rho\ =\ \frac{1}{2}\frac{\tilde{\pi}_T^2}{\det q}\ =\ T_{\mu\nu}n^\mu n^\nu
\end{equation}
is the \emph{energy density of the scalar field with} respect to the
class of observers comoving with the universe.}

{The quantum energy density operator and its spectrum is another
  subject discussed in this paper on its own.  The operator is used in
  the APS model as the measure of the avoidance of
the singularity.  At the early stages of LQC it was believed that the
singularity avoidance is a kinematical effect implied by the
non-singular way the metric determinant inverse shows up in the
expression of the energy density. Indeed, the LQG motivated
quantization of that expression has (up to factor ordering ambiguity)
the form
\begin{equation}
  \hat{\rho}\ =\ \frac{1}{2}\hpp^2 \widehat{\det q^{-1}} \ ,
\end{equation}
where the operator $\widehat{\det q^{-1}}$ is bounded, and actually
annihilates  the vector annihilated  by $\widehat{\det q}$. A
stronger result takes place in the APS model. Namely, the
expectation value of the energy density
$\langle\hat{\rho}\rangle(T)$ evolving with the time $T$ approaches
certain universal value (of the order of Planck energy density
$\rhoPl$){\footnote{Throughout of
    this paper we use the value of $\rhoc$ derived in
    \cite{aps-imp}. However recently it was shown \cite{entropy} that
    due to subtleties in constructing the loop of minimal area in LQC
    the so called area gap (lowest nonzero area eigenvalue) is twice
    bigger than the one used in \cite{aps-imp}. In consequence the
    value of $\rhoc$ (depending on it) is twice smaller and equals
    approximately $0.41\rhoPl$.}}
\begin{equation}
  \langle\hat{\rho}\rangle(T)\ \leq\ \rhoc\
  \approx\ 0.82\rhoPl
\end{equation}
from below, and bounces back. Here we show, that the essential
spectrum of $\hat{\rho}$ is
\begin{equation}
  {\Sp{}_{\ess}}(\hat{\rho})\ =\ [0,\rhoc] \ .
\end{equation}
There may still exist discrete spectrum elements bigger then
$\rhoc$, however, the corresponding eigenfunctions are focused near
the zero volume and therefore their contribution to semiclassical
states focused at large {scalar field momentum} (and so at large
volumes) is extremely small.}

\subsection{{The role of the zero volume state}}
\label{sec:out-zero}

{A technical subtlety concerning the constraint operators  above,
is that in the APS model the zero volume state $|0\rangle \in
{\cal H}_{\rm gr}$ is at
the same time annihilated by the inverse-volume operator
\begin{equation}
  \widehat{\sqrt{\det q}^{-1}}|0\rangle\ =\ 0 \ .
\end{equation}
This leads to {an impression of} incompleteness in a definition
of the operator $\Ctot(N')$ in
$(\Hgr,\,(\cdot|\widehat{\sqrt{\det q}^{-1}}\cdot))$ present even
after the modification of the scalar product which removes that zero
volume state.  The solution to that {subtlety} is hidden in the
results published in the literature
\cite{aps-imp,apsv-spher,negL}, but it  {has never been} spelled
out. We will present the details in Sec. \ref{sec:zero} showing in
particular  some constraint  {being} induced in $\Hgr$ by the scalar
constraint operator. The presence of this constraint allowed to
define {rigorously} the evolution operator in \cite{aps-imp} and
following works.} {The discussed structure allows in
particular to immediately extend the results of \cite{kl-sadj} to
superselection sector containing the $|v=0\rangle$ state.}

\section{The elements of the LQC FRW}
\label{sec:LQC}

In LQC, like in the other cosmological models, one restricts the
Einstein's theory to the space of the space-time metrics and other
fields having a given symmetry. Here we consider the case of
Friedman-Robertson-Walker (FRW) models corresponding to homogeneous
and isotropic spacetimes. In this section we briefly introduce the
quantum description of these models within LQC framework. For
shortness we will introduce only those elements of the LQC models
which will be relevant for our studies. For more detailed description
of the quantization procedure the reader is referred to \cite{abl} and
\cite{aps-imp}.

On the classical level the spacetime is described by the product
manifold $\mathbb{R}\times\Sigma$ and a metric tensor
\begin{equation}
  \rd s^2\ =\ -N^2 \rd t^2\ +\ a^2\fidq_{ab}\rd x^a\rd x^b
\end{equation}
where $\fidq$ is a fixed, auxiliary, homogeneous, isotropic metric
tensor on $\Sigma$, and $N$ is a homogeneous lapse function.  The
metric is coupled with a scalar field $T$ homogeneous on $\Sigma$.
These properties boil down to conditions
\begin{subequations}\label{eq:afhom}\begin{align}
  a(t,x)\ &=\ a(t) \ ,  &
  T(t,x)\ &=\ T(t) \ ,  &
  N(t,x)\ &=\ N(t) \ .
\tag{\ref{eq:afhom}}\end{align}\end{subequations}

The diffeomorphism constraints are trivially satisfied, hence the
only Einstein constraint is the scalar constraint. It takes the
following form
\begin{equation}
  \Ctot(N) \ =\ N(\Cgr + \frac{1}{2}\frac{\tilde{\pi}_T^2}{|V|}) \ ,
\end{equation}
where  one fixes a finite region ({``cell''})
${\Sigma}_0\subset\Sigma$ to integrate (if $\Sigma$ is compact a
natural choice is $\Sigma_0=\Sigma$)
\begin{subequations}\label{eq:basic-int}\begin{align}
  \pp\ &:=\ \int_{\Sigma_0}{\tilde{\pi}}_T \ ,  &
  |V|\ &:=\ \int_{\Sigma_0} a^3\sqrt{{\rm det}q^{(0)}} \ ,  &
  C_{\rm gr}\ &=\ \int_{\Sigma_0}\tilde{C}_{\rm gr}
\tag{\ref{eq:basic-int}}\end{align}\end{subequations}
and $\tilde{C}_{\rm gr}$ is the Hamiltonian density of the
gravitational field. One also introduces the oriented volume function
ranging from $-\infty$ to $\infty$, namely
\begin{equation}
  V\ =\ \pm|V| \ ,
\end{equation}
with the sign depending on the orientation in $\Sigma_0$ of the
triad with respect to a fixed fiducial orientation of $\Sigma$. The
kinematical Hilbert space and the quantum operators of the scalar
field $T$ and its conjugate momentum $\pp$ are
\begin{subequations}\begin{align}
  {\cal H}_{\rm sc}\ &=\ L^2(\R) \ , \\
  \hat{T}\psi(T)\ =\ T\psi(T) \ ,\qquad  &{}\qquad
  \hpp\psi(T)\ =\ -i\hbar\partial_T\psi(T) \ .
\end{align}\end{subequations}
The kinematical Hilbert space and the basic quantum operators for the
gravitational field in the APS and sLQC model are,
\begin{subequations}\begin{align}
    \Hgr\ &=\ \overline{\Span(|v\rangle\ :\ v\in\R)} \ ,  &
    \langle v|v'\rangle\ &=\ \delta_{v,v'} \ , \\
    \hat{V}|v\rangle\ &=\ \left(\frac{8\pi\gamma}{6}\right)^{\frac{3}{2}}
      \frac{3\sqrt{3\sqrt{3}}}{2\sqrt{2}}\,v\,\lPl^3|v\rangle\
        =:\ V_o\, v|v\rangle \ ,  &
    \hat{h}_{\nu}|v\rangle\ &=\ |v+\nu\rangle \ ,
\end{align}\end{subequations}
where the operator $\hat{h}_{\nu}$ is a shift operator -- a component
of an operator corresponding to the classical holonomy function
involving $d a/dt$.

The kinematical Hilbert space of the system is the tensor product
$\Hsc\otimes\Hgr$. Every element $\psi\in \Hsc\otimes\Hgr$
is thought of as a function of the variables $T$ and $v$, and its
values will be denoted by $\psi(T,v)$.

The quantum scalar constraint is considered in the following form
\begin{equation}\label{c}
  \left(\frac{1}{2}\hpp^2\otimes 1\ +\ 1\otimes
  \widehat{|V|^{-1}}^{-1}\widehat{C_{\rm gr}}\right)\,\psi(T,v)\
  =\ 0 \ ,
\end{equation}
where:
\begin{itemize}
  \item $\widehat{|V|^{-1}}=V_o^{-1}B(\hat{V})$
    is a result of a quantization of the classical $1/|V|$, with $B$
    being a function. In the orthodox LQC it descends from the LQG
    definition of the orthonormal coframe expressed by commutators of
    various powers of the volume operator. For the studies performed
    in this article the exact form of $B$ does not matter. What is
    important are the following properties (true for both APS LQC and
    sLQC)
    \begin{enumerate}[(i)]
      \item $B(v) = B(-v)$,
      \item {for non zero $v$} is finite and nonvanishing, and
      \item for large $|v|$,  $B(v) \simeq \frac{1}{|v|}$.
    \end{enumerate}
    More specific assumptions will be made whenever necessary.
    Particular form of $B$ in models considered here is, respectively,
    \begin{subequations}\label{eq:B-form}\begin{align}
      B_{\rm sLQC}(v)\ &=\ \frac{1}{|v|} \ ,  &
      B_{\rm APS}(v)\ &=\ \frac{27}{8}|v|
      \big||v+1|^{\frac{1}{3}}-{|v-1|}^{\frac{1}{3}}\big|^3
      \ .
    \tag{\ref{eq:B-form}}\end{align}\end{subequations}
  \item the operator $\widehat{C_{\rm gr}}$ has the form
    \begin{equation}\label{cgr}
      \widehat{C_{\rm gr}}\ =\ i(\hat{h}_{2} -
      \hat{h}_{-2})A(\hat{V})i(\hat{h}_{2}-\hat{h}_{-2})
      - \Lambda\hat{V} + W_k(\hat{V})
    \end{equation}
    with $\Lambda$ being the cosmological constant, and $A$, $W_k$ being
    suitable symmetric functions, the second one depending on the type of the
    local symmetry group ($k=-1,0,1$). The  assumption about $A$ we
    will refer to is the behavior $A(v)\sim |v|$ for large $|v|$ true in
    LQC as well as in the sLQC. In these two particular
    cases the form of $A$ reads
    \begin{subequations}\label{eq:A-form}\begin{align}
      A_{\rm APS}\ &=\ A_o\tilde{A}\ =\  A_0|v|\big||v+1|-|v-1|\big|
      \ , & & \\
      A_{\rm sLQC}\ &=\ 2A_o |v| \ ,  &
      A_o\ &:=\ \frac{9\sqrt{3}\lPl}{32\sqrt{\pi}\gamma^{\frac{3}{2}}G}
      \ .
    \end{align}\end{subequations}
\end{itemize}

The physical states are solutions to the quantum constraint,
according to the APS model, thought of as maps
\begin{equation}\label{c'}
  \mathbb{R} \ni T\ \mapsto\ \psi_T \in \Hgrb \ ,
\end{equation}
where the space ${\cal H}_{\rm gr, B}$ {(referred to further as
  an \emph{auxiliary} space)} is defined by the same ${\rm
Span}(|v\rangle\! :\ v\in\R)$ as before, however endowed by APS with
the scalar product
\begin{equation}
  (\cdot|\cdot)_B\ =\ \langle \cdot |B(\hat{V})\cdot\rangle \ .
\end{equation}
That definition of the new scalar product is suited to make the {\it
evolution operator}
\begin{equation}\label{eq:Theta}
  \hat{\Theta}\ :=\ - (B(\hat{V}))^{-1}\Cgr
\end{equation}
symmetric, however the definition of this operator {in the
form it
  is presented above,} needs to be completed. {Such precise
  definition, which was
used in \cite{aps-imp,apsv-spher,negL}, is discussed in Appendix
\ref{sec:zero}.} Now, each solution to the scalar
constraint takes the form
\begin{equation}\label{decomp}
  \psi\ =\ \psi^{-}\ +\ \psi^+ \ ,
\end{equation}
where $\psi^{\pm}$  satisfies, respectively,
\begin{equation}\label{cpm}
  \hpp\psi^{\pm}_T(v) \ =\ \pm
  \sqrt{2V_o\hat{\Theta}}\, \psi^{\pm}_T(v) \ ,
\end{equation}
where each solution $\psi$ of (\ref{cpm}) takes values $\psi_T$  in
the part of the Hilbert space corresponding to the non-negative part
of the spectrum of the operator $\hat{\Theta}$, and the square root
is defined on that subspace. We will be assuming throughout this
paper that this decomposition is unique, {which} is generically
true\footnote{That is as long, as $0$ is not an eigenvalue of
  $\hat{\Theta}$.}. A non-unique case is considered in \cite{posL1}.
Given two solutions $\psi$ and $\psi'$ of the quantum scalar
constraint, APS define the following scalar product
\begin{equation}\label{scprphys}
  (\psi|\psi')_{\rm phys}\ :=\ (\psi^+_T|\psi^+_T)_B\ +\
  (\psi^-_T|\psi^-_T)_B
\end{equation}
where the RHS is independent of $T$. Denote the resulting Hilbert
space by $\Hphys$.

A physical observable $\hpp$ is
\begin{equation}\label{pi}
  \hpp\psi^\pm\ :=\ \pm
  \sqrt{2V_o\hat{\Theta}}\, \psi^\pm \ ,
\end{equation}
The volume operator $\hat{V}$  defined in the kinematical Hilbert
space gives rise to the physical observable ${\hat{\cal
O}}_{V}(T_0)$ (modulo the discussion in Sec. \ref{sec:phys-oper}
below) determined by a number $T_0$ (the ``instant of time'') and
defined by the following {expression}
\begin{equation}
  ({\hat{\cal O}}_{V}(T_0)\psi)_{T_0}\ =\ \hat{V}\psi_{T_0} \ .
\end{equation}
In consequence it can be thought of as an operator in QM acting at an
instant $T_0$ on a state evolving in the Schroedinger picture.

The final Hilbert space is selected as the irreducible subspace of all
the quantum observables we choose. Classically, the system can be
described completely by scalar field momentum $\pp$ and the volume of
the fixed cell $V$. The first one commutes with the constraint,
whereas the second defines the Dirac observables via the relational
observables construction. Therefore, APS assume that the sufficient
set of quantum operators to describe every quantum state consists of
the following operators:
\begin{subequations}\label{eq:obses}\begin{align}
  \hpp\psi^{\pm}(T,v)\
  &:=\ (\id\times\sqrt{2V_o\hat{\Theta}})\,\psi^{\pm}(T,v) \ ,  \\
  |\hat{V}|_{T_0}\psi^{\pm}(T,v)\
  &:=\ e^{\pm i(T-T_0)\sqrt{2V_o\hat{\Theta}}}\,\hat{V}\,\psi^{\pm}(T_0,v) \ .
\end{align}\end{subequations}
{Furthermore}, there are subspaces  $\Hil^\pm_{{\phys},\epsilon}$ preserved by the
action of all the quantum observables, labeled by arbitrary
$\epsilon\in [0,4)$ and a  sign `$+$' or `$-$'  corresponding to the
decomposition (\ref{decomp}).  The subspace ${\cal H}^+_{{\rm
phys},\epsilon}$ (${\cal H}^-_{{\rm phys},\epsilon}$)  is the space
of solutions (\ref{c'}) to (\ref{cpm}) with $\pm=+$ ($\pm=-$) which
take values in the following subspace of ${\cal H}_{{\rm gr},B}$
\begin{equation}\label{Hpmepsilon}
  {\cal H}_{\epsilon}\ =\ \overline{\Span(|v\rangle\ :\
    v=\epsilon+4n,\ \ n\in\Z)}
\end{equation}
In the special cases of $\epsilon=0,\,2$, the subspace ${\cal
H}^\pm_{\epsilon}$ admits the action of the orientation changing
operator
\begin{equation}
  (P\psi)_T(v)\ \mapsto\ \psi_T(-v) \ .
\end{equation}
which commutes with $T$. In that case APS restrict the Hilbert space
${\cal H}^\pm_{\epsilon}$ further, to the subspace of the even
functions.

The operator $\hat{\Theta}$ is well defined in every subspace
$\Hil^\pm_{\epsilon}$ (in the domain ${\Span(|v\rangle\ :\
v=\epsilon+4n,\ \ n\in\Z)}$) such that $\epsilon\not=0$, however for
$\epsilon=0$ its definition is not {\it a priori} obvious and needs
explanation.  We provide it in {Appendix \ref{sec:zero}} as well as our
definition of the evolution operator $\hat{\Theta}$ in the $B(0)=0$
case.
\medskip

\noindent{\bf Remarks.} \begin{itemize}
  \item For the remaining values of the parameter $\epsilon$ we have
    $P(\Hil^\pm_{\epsilon})=\Hil^\pm_{4-\epsilon}$. Then, APS construct
    the space of the even functions spanned by elements of
    $\Hil^\pm_{\epsilon}$ and $P(\Hil^\pm_{\epsilon})$. In these cases
    construction reduces (is unitarily equivalent) to the single
    $\Hil^\pm_{\epsilon}$.
  \item We will often ignore the reducibility and consider the
    whole Hilbert space $\Hil_{\phys}$.
\end{itemize}

\section{The quantum geometry operators in the physical theory}
\label{sec:phys-oper}

In the Dirac program one of the most common techniques of
constructing observables on $\Hphys$ is an appropriate pull-back
onto it of kinematical ones. However unless the quantity measured by
given observable is a constant of motion such direct pull-back will
not correspond to any physically interesting property of the system.
Therefore in such cases one tries to construct operators measuring
kinematical quantity ``at a given time'' (example of which is the
operator $|\hat{v}|_{\phi}$ defined in \cite{aps-imp}). Technically
this corresponds to the pull-back of kinematical observable to an
auxiliary Hilbert space, the image of mapping \eqref{c'}. In this
section we address (in context of LQC) the question of how the
original properties of kinematical operators transfer to physical
spaces. We will see that even the volume operator, seemingly under a
perfect control, may surprisingly change much more, than it is
expected in the LQC  literature.

{Let us start our analysis in the context of an APS
  LQC, where $B(0)=0$. There the auxiliary Hilbert space is a space
  $\Hil_{\gr,B}$ equipped with}
the modified scalar product $\langle\cdot|B(\hat{V})\cdot\rangle$.
The quantum volume operator is unchanged by this modification, and
is still well defined and essentially self-adjoint in the domain
$\Span(|v\rangle\ :\ v\in\mathbb{R})$. However a general operator
$\hat{g}$ defined in that domain in ${\cal H}_{\rm gr}$ should be
redefined such that modulo the ordering it corresponds to the same
classical kinematical observable, but has the correct properties
with respect to the ${}^\dagger$ operation. An example of such
redefinition is replacing $\hat{g}$ defined in ${\cal H}_{\rm gr}$
by
$$B(\hat{V})^{-1/2}\hat{g}B(\hat{V})^{1/2}$$
defined in ${\cal H}_{{\rm gr},B}$. In fact, this transformation
coincides with the pull back of $\hat{g}$ by the unitary map used in
the previous section, {which is}
\begin{subequations}\label{eq:tr2}\begin{align}
    \Hil_{\gr,B}\ &\rightarrow\ \Hgr  \ ,  &
    \psi\ &\mapsto\ \sqrt{B(\hat{V})}\psi \ ,
\tag{\ref{eq:tr2}}\end{align}\end{subequations} and the inverse
image of the domain $\Span(|v\rangle\ :\ v\in\mathbb{R})$ is
$\Span(|v\rangle\ :\ 0\not= v\in\mathbb{R})$. As we mentioned, the
volume operator $\hat{V}$ is not affected by that transformation
(modulo the small restriction of the domain  {consisting in}
disappearing of the zero volume eigenvector $|0\rangle$.)

In the sLQC case, on the other hand, {the auxiliary space
  is directly}
$\Hgr$, so {the analog of the transformation \eqref{eq:tr2} is just
an identity}.

{The transformation presented above does not however solve
  all the problems. To see that} let us go back to the construction of
the physical Hilbert space.
{To shorten the explanation we introduce the common notation
denoting} by $({\cal H},(\cdot|\cdot))$ the Hilbert space $({\cal
H}_{{\rm gr},B},\langle\cdot|B\cdot\rangle)$ in the APS model case,
or $({\cal H}_{{\rm gr}},\langle\cdot|\cdot\rangle)$ in the sLQC
case. {Then each element of $\Hphys$ is represented by a mapping}
\begin{equation}
  \R\ni T\mapsto\psi_T\in{\cal H}
\end{equation}
where the ${\cal H}$ valued functions $\psi$ satisfy the equation
\begin{equation}\label{thec}
  {\frac{\partial^2}{\partial T^2}}\psi_T\ =\
  -2V_o\hat{\Theta}\psi_T \ .
\end{equation}
Choosing any instant of $T=T_0$, we have two maps from the space of
the solutions to $\Hil$,
\begin{equation}\label{themap}
  \psi\mapsto \psi^{\pm}_{T_0}\in {\cal H} \ ,
\end{equation}
{corresponding, respectively, to the positive and negative
  frequency solutions.}
Let us fix one of them (that is either `$+$' or `$-$'). {Now}, the
important observation is, that if the operator $\hat{\Theta}$ is not
positive {(which happens for
  example in the case $\Lambda>0$)}, then the image of the map
(\ref{themap}) is not the entire Hilbert space ${\cal H}$. Indeed, a
physical solution should satisfy at every instant $T_0$,
\begin{equation}\label{eq:positive}
  (\psi_{T_0}\,|\hat{\Theta}\psi_{T_0})\ \ge\ 0 \ .
\end{equation}
Assuming that the operator $\hat{\Theta}$ is self-adjoint, we can
identify the image of the map with the subspace
$\Hil_{\hat{\Theta}\geq0}$ of $\Hil$ corresponding to the non-negative part
of the spectrum of $\hat{\Theta}$.

Let us now consider an example of the operator $\hat{g}$, the volume
$\hat{g}\ =\ \hat{V}$. For the pullback of the operator at any $T$
to ${\cal H}_{\rm phys}$ to be well defined, the answer to the
following two questions should be affirmative:
\begin{enumerate}[(i)]
  \item \label{it:q1} Is any dense subset of $\Hil_{\hat{\Theta}\geq 0}$
    contained in the (maximally extended) domain of the operator
    $\hat{V}$?
  \item \label{it:q2} Is the space $\Hil_{\hat{\Theta}\geq 0}$ preserved
    by the volume operator $\hat{V}$?
\end{enumerate}

When $\Lambda >0$,  the answer to the question \eqref{it:q1} is
likely to be  negative. In particular, we do know that the
eigenvectors of the evolutions operator $\hat{\Theta}$ are not in
the domain of the volume operator.  {A heuristic reason for that can
be seen at
  the classical level already, when the trajectories reach infinite
  volume for finite $T$. To avoid this problem one has to
  ''compactify'' the volume, that is to consider, instead of
  $\hat{V}$, an operator $f(\hat{V})$, where $f$ is a bounded (but
  monotonic) function.}

The most likely answer to the question \eqref{it:q2} is also ``no''.
{This means that}, given a solution $\psi$ of (\ref{thec}) at an
instant $T_0$ (taking values in the subspace $\Hil_{\hat{\Theta}\geq
0}$), in general there is no solution $\psi'$ of (\ref{thec})  such
that
\begin{equation}
  f(\hat{V})\psi_{T_0}\ =\ \psi'_{T_0} \ .
\end{equation}
{To overcome this problem} one can employ the fact, that  the
sesquilinear form $(\cdot|f(\hat{V})\cdot)_B$ defined by
$f(\hat{V})$ can be restricted to any subspace and define an
operator therein. This is equivalent to using the orthogonal
projection
\begin{equation}\label{eq:pos-proj}
  P_{\Theta\geq 0}:\Hil_{{\rm gr},B}\rightarrow \Hil_{\Theta\geq 0}
\end{equation}
and replacing he operator  $f(\hat{V})$ by
\begin{equation}
  f(\hat{V})'\ :=\ P_{\Theta\geq 0}f(\hat{V})P_{\Theta\geq 0}.
\end{equation}
The final operator $f(\hat{V})'$ is a well defined
  observable, in a sense that the answer to both \eqref{it:q1} and
  \eqref{it:q2} is  affirmative.

To summarize, in the case when $\Theta$ is not positive definite the
straightforward pull-back of the kinematical volume operator does
not define correct physical observable. To define it correctly one
has to implement additional modifications, like the ones presented
above. {However, one} should be aware of the likelyhood of change of
the commutation relations between projected operators, as in general
for a projection operator $P$ we have
\begin{equation}
  [PAP,PBP]\ \not=\ P[A,B]P \ .
\end{equation}

Finally, {let us} consider a relation of the physical volume
operator with the original, kinematical one in the APS model.

When the operator $\hat{\Theta}$ is positive, the map
\begin{equation}\label{iso}
  \Hil^\pm_{\phys}\ni\psi\mapsto \psi_{T_0}\in
  \Hil_{\gr,B}
\end{equation}
is unitary. It pulls back the operator $\hat{V}$  to the observable
operator $\hat{{\cal O}}_{V}(T_0)$. Hence, the spectrum of the
resulting physical  operator observable $\hat{{\cal O}}_{V}(T_0)$
coincides with the spectrum of the restriction of $\hat{V}$ and it
is independent of $T_0$.

{The} situation changes if the operator $\hat{\Theta}$ is
non-definite. The
physical operator is now the pullback by (\ref{iso}) of the modified
operator $P_{\Theta\geq 0}f(\hat{V})P_{\Theta\geq 0}$ which is just a
different operator than $f(\hat{V})$. In consequence their spectra
may differ considerably.

\section{The spacetime metric tensor operator from LQC}
\label{sec:qmetric}

Having the LQC FRW model at our disposal, we can implement our
consideration from Sec. \ref{sec:pre-qsp} concerning a quantum
spacetime metric operator. For this model the classical spacetime
metric tensor is
\begin{equation}
  \rd s^2\ =\ -\frac{V^2}{\pp^2}\rd T^2\ +\
\frac{{V}^{\frac{2}{3}}}{\int_{{\cal U}_0}\sqrt{{\rm det}\fidq}}
  \,{\fidq_{ab}\rd x^a\rd x^b} \ .
\end{equation}
Applying the discussion of Sec. \ref{sec:pre-qsp} regarding lapse
function we observe that the quantum operator corresponding to $\rd
s^2$ should have the form
\begin{equation}
  -\widehat{{\left(\frac{V^2}{\pp^2}\right)}}\rd T^2\ +\
  \frac{{\hat{V}}^{\frac{2}{3}}}{\int_{{\cal U}_0}\sqrt{{\rm
        det}\fidq}}
  \,{\fidq_{ab}\rd x^a\rd x^b}
\end{equation}
where ${\widehat{\left(\frac{V^2}{\pp^2}\right)}}$ stands for a
quantum operator
corresponding to the classical  ${\frac{V^2}{\pp^2}}$. However, the
operators $\hat{V}$ and $\hpp$ do not commute in ${\cal H}_{\rm
phys}$ (see (\ref{pi})). Therefore there are two possibilities:
\begin{enumerate}[(i)]
  \item the time part of the space time metric is only a semiclassical
    notion, and does not exists as a uniquely defined quantum operator,
    or
  \item physics chooses one specific way of defining that operator,
    however we do not have sufficient information to guess that choice.
\end{enumerate}

Remarkably, however, quantum test fields interacting with this quantum
spacetime propagate in the unique way independent of that ambiguity
\cite{qft-qsp}. Thus, the possible physical answer to that issue may
be that the quantum metric is defined uniquely only through matter
propagating on it.

The spacetime metric tensor, if it exists, can be used to calculate
the geometric time of an interval $((T_1,x^a),(T_2,x^a))$ in a state
(\ref{c'}). It is given by the following formula
\begin{equation}
  \tau_{T_2,\,T_1}\ =\ \int_{T_1}^{T_2} {\sqrt{(\,\psi_T\,
    |\,\widehat{\left(\frac{V^2}{\pp^2}\right)} \psi_{T}\,)}\rd T.}
\end{equation}

Classically, the time component of the metric tensor can be
expressed by the energy density ${\rho}$ of the homogeneous scalar
field. The relation reads
$$  {\frac{V^2}{\pp^2}}\rd T^2\ =\ {2}{\rho}^{-1}\rd T^2 $$
Assuming that the relation holds on the quantum level, we have
\begin{equation}\label{eq:rho-appear}
  {\widehat{\left(\frac{V^2}{\pp^2}\right)}}\rd T^2\ =\
  {2}\hat{\rho}^{-1}\rd T^2.
\end{equation}
However, we still have the similar ordering freedom in the
definition of $\hat{\rho}$ operator {(see section
\ref{sec:rho})}.

\section{Non-equivalence of the constraints $\widehat{C(1)}$ and
$\widehat{C(|V|)}$} \label{sec:C-noneq}

In the previous sections, following the APS approach, we considered
the scalar constraint in the form (\ref{c}), that is
\begin{equation}\label{c1}
\widehat{C(|V|)}\ =\ \left(\frac{1}{2}\hpp^2\otimes 1\ -\
  1\otimes V_o\hat{\Theta}\right)
\end{equation}
defined in the Hilbert space ${\cal H}_{\rm sc}\otimes{\cal H}_{{\rm
gr},B}$ (see Sec. \ref{sec:LQC}, App. \ref{sec:zero}).

This quantum  constraint  corresponds to the classical scalar
constraint $C(N_1)$ given by the choice of the lapse function
\begin{equation}\label{N1}
  N_1\ =\ |V| \ .
\end{equation}

On the other hand one can choose different lapse, {in particular}
\begin{equation}\label{N2}
   N_2\ =\ 1 \ ,
\end{equation}
{which is more natural from the point of view of full LQG.} The
corresponding quantum constraint operator is {of the form}
\begin{equation}\label{BC}
  \widehat{C(1)}\ =\  ( \frac{1}{2}\hpp^2\otimes
  |\widehat{V^{-1}}|
  + 1\otimes \widehat{C_{\rm gr}})
\end{equation}
and it is defined right in the kinematical Hilbert space
$\Hsc\otimes\Hgr$.

Given the quantum scalar constraint operator in either of the forms,
the general construction (via the method of group averaging
  \cite{gave}) of the physical Hilbert space  uses its
spectral decomposition. The solutions are distributions defined on
the spectrum and supported at the point $\lambda=0$. In the case of
the operator (\ref{c1}) the construction boils down to the APS
construction outlined in Sec. \ref{sec:LQC} \cite{aps}. {In this
section we address the question} whether the second choice of the
lapse function {\eqref{N2}} leads to the same result.

{To find an answer to this question we have to compare the spectral
properties of constraints \eqref{c1} and \eqref{BC}. In the first
case they are encoded in properties of the family of operators
$\Theta_{\pp} := \frac{1}{2}\pp^2-V_o\hat{\Theta}$, (with $\pp\in
\mathbb{R}$) depending in turn on the spectral structure of
$\hat{\Theta}$ in ${\cal H}_{{\rm gr},B}$.
In the second case, on the other hand, the constraint inherits its
properties from the family $\widehat{C_{\gr,\pp}} :=
\frac{1}{2}\pp^2 |\widehat{V^{-1}}|+{\Cgr}$ (with
$\pp\in\mathbb{R}$) defined in $\Hgr$. In both cases the domain of
considered families is ${\rm Span}(|v\rangle\ :\ v\in \mathbb{R})$}

Let us first turn our attention to the case of constraint
\eqref{c1}. As discussed above its properties (in particular
self-adjointness) are inherited from $\hat{\Theta}$, which has been
recently  extensively investigated. In particular
\begin{enumerate}[(i)]
  \item for
    \begin{itemize}
      \item $\Lambda <0,\, k=-1,0,1$,
      \item $\Lambda=0,\, k=0, k=1$
      \item $\Lambda > \Lambda_{\crit},\, k=-1,0,1$,
      \item $\Lambda = \Lambda_{\crit},\, k=0$
    \end{itemize}
    (where $\Lambda_{\crit}:=8\pi G\rhoc$) the operator $\Theta$ is
    essentially self-adjoint \cite{kl-sadj,sa}, whereas
  \item for $\Lambda\in (0,\Lambda_{\crit}), k=0$ {\emph{it has
      inequivalent self-adjoint extensions} (see, for detailed
    analysis, \cite{posL1,posL2} and also \cite{sa} for a summary, all
    currently in preparation)}.
\end{enumerate}

{In the latter case} each self-adjoint extension of the operator
$\hat{\Theta}$ defines via the APS construction a distinct quantum
theory. The unitary non-equivalence of the different extensions
follows from the difference between  the corresponding discrete
spectra.

That non-uniqueness in the self-adjoint extensions is related to the
properties of the classical system: the evolution of the FRW
spacetime reaches the end (the infinite physical time of the
observers expanding with the universe {also corresponding to an
infinite volume}) in a finite value of the scalar field $T$ used as
a time variable. {In consequence to continue evolution in $T$ one
has to specify boundary conditions at $|v|=\infty$.}
\medskip

Surprisingly, those properties of the constraint operator (\ref{c1})
in the case $\Lambda>0$, are in  contrast with the properties of the
constraint operator (\ref{BC}) corresponding to the choice of the
lapse function $N_2 = 1$, namely:
\begin{obs}\label{thm:o1} {The operator
  $\widehat{C_{\rm gr}}$ is essentially self adjoint for arbitrary
  value of the cosmological constant $\Lambda$ and for arbitrary case
  $k=-1,0,1$.}
\end{obs}
The technical reason {for this} is the following general fact
{\cite{jacobi}}:
\begin{lem}\label{thm:l1}
  In the Hilbert space
  \begin{equation}
    \overline{\Span(|v\rangle\ :\ v\in
      \mathbb{R})}\ \ \ \langle v|v'\rangle\ =\ \delta_{v,v'}
  \end{equation}
  consider an operator
  \begin{equation}
    (h_{2}-h_{-2})\tilde{A}(\hat{V})(h_{2}-h_{-2}) + W(\hat{V})
  \end{equation}
  defined in the domain ${\Span(|v\rangle\,:\,v\in\mathbb{R})}$. That
  operator is essentially self-adjoint for every function $W$ and
  every nowhere-vanishing function $\tilde{A}$ such that
  \begin{equation}\label{sum1/A}
    \sum_{n\in\mathbb{Z}_{+}}\frac{1}{|\tilde{A}(\epsilon+4n)|}\ =\ \infty\quad
    {\rm and}\quad
    \sum_{n\in\mathbb{Z}_{-}}\frac{1}{|\tilde{A}(\epsilon+4n)|}\ =\ \infty \ .
  \end{equation}
\end{lem}

Note, that the result holds for $\tilde{A}=A$ due to the {asymptotic
behavior} $A(v)\propto |v|$ for $|v|\rightarrow \infty$. {On the other
hand it does not hold for constraint (\ref{c1}) for} considering it
amounts to replacing the function $A$ by a function $\breve{A}\propto
|v|^2$ {(see \eqref{c})} which does not satisfy the condition
(\ref{sum1/A}).
\medskip

The remaining {(not covered by Obs.\ref{thm:o1})} term in \eqref{BC}
\begin{equation}
  \frac{1}{2}\pp^2 \widehat{V^{-1}}\ =\ \frac{1}{2V_o}\pp^2 B(\hat{V})
\end{equation}
is bounded (in the APS case) and does not spoil the essential-self
adjointness while added to $\widehat{C_{\rm gr}}$. {As the
consequence,} self-adjoint extensions of the constraint operator
\eqref{BC} is uniquely defined.
\medskip

In summary, we have considered two operators (\ref{c1}) and
(\ref{BC}) of the quantum scalar constraint corresponding to the
classical constraint $C(N)$ with two different choices of a lapse
function, namely: $N=N_1$ (\ref{N1}) and $N=N_2$ (\ref{N2}),
respectively. {Provided,} the cosmological constant $\Lambda<0$,
each of the operators $\widehat{C(|V|)}$ and $\widehat{C(1)}$
has a uniquely defined self-adjoint extension.  However, if the
value of the cosmological constant is positive $\Lambda_{\rm
cr}>\Lambda>0$, then the quantum scalar constraint operator
$\widehat{C(|V|)}$ depends on a choice of a self adjoint
extension of the operator $\hat{\Theta}$. Each choice determines a
{(potentially) distinct} physical model. The operator
$\widehat{C(1)}$ on the other hand, is essentially self-adjoint for
every value of $\Lambda$.

How do those results fit together? What is the comparison between
the sets of solutions to the quantum scalar constraint  defined by
using the operator $\widehat{C(|V|)}$ as opposed to those
defined by using the operator $\widehat{C(1)}$?

It turns out, that in every case with $\Lambda<0$, the solutions to
the quantum scalar constraint $\widehat{C(|V|)}$ coincide  with
the solutions to the quantum scalar constraint $\widehat{C(1)}$ and
the physical model is independent of which constraint operator we
use to construct it.

Let us turn now to the {positive} $\Lambda>0$ case. {One can ask:}
what are the physical solutions obtained from the spectral
decomposition of the operator $\widehat{C(1)}$. To answer it, let us
restrict (for simplicity) the space $\Hgr$ to the subspace
\begin{equation}
  \Hil_{o}^{\rm ev}\ :=\ \Span(|v\rangle + |-v\rangle\,:\,0\not=
  v\in\mathbb{R} ) \ .
\end{equation}
That subspace is preserved by the operator ${\Cgr}$, and actually,
is exactly the subspace promoted to the physical Hilbert space in
\cite{aps}, which makes our restriction justified. A physical
solution $\psi$ obtained by the spectral decomposition of
$\widehat{C(1)}$ restricted to $\Hil_{o}^{\rm ev}$ is a family
\begin{equation}
  [0,\pi)\ni a\mapsto \psi^{(a)}
\end{equation}
of solutions to the constraints
\begin{equation}
  \widehat{C(|V|)}^{(a)}\psi^{(a)}\ =\ 0 \ ,
\end{equation}
where the $a$ labels the self-adjoint extensions of the constraint
operator, and $\widehat{C(|V|)}^{(a)}$ stands for the
corresponding extension. The physical scalar product derived from
the spectral decomposition of $\widehat{C(1)}$ is
\begin{equation}
  (\psi|\psi')\ =\
  \int_{0}^\pi \rd a {(\psi^{(a)}|\psi'^{(a)})}_{\rm phys} \ ,
\end{equation}
where ${(\psi^{(a)}|\psi'^{(a)})}_{\phys}$ is the physical scalar
product \eqref{scprphys} between the states in the APS model.

In conclusion, the physical Hilbert space constructed directly from
the constraint \eqref{BC} ``contains'' {\it all} the solutions given
by {\it all the self adjoint extensions} of the operator {$\Theta$}.
The reason {for}  the quotation mark is that the solutions to
\eqref{c1} are not normalizable solutions of \eqref{BC}. The
detailed construction of the Hilbert space of solutions to the
constraint $\widehat{C(1)}$ will be presented  in \cite{ga}.

\section{The scalar field energy density operators}
\label{sec:rho}

{Consider now the quantum scalar field energy density operator
$\hat{\rho}$ introduced in \eqref{eq:rho-appear}. For the flat
isotropic FRW systems considered so far in LQC \cite{aps-imp,negL} the
analysis of states semiclassical at late times has shown that the
expectation values of $\hat{\rho}$ for such states are always bounded
from above by a fundamental value $\rho_c$. This result was next
extended in context of sLQC (for $k=0$, $\Lambda=0$) to full physical
Hilbert space \cite{sLQC}. In this section we address the issue of the
boundedness of energy density both in solvable and APS formulation
(see Sect. \ref{sec:LQC}) of LQC from a slightly different perspective,
namely by analysing the spectrum of the operator $\hat{\rho}$.}

{Here,} for the convenience, we will work with a slightly different
representation of the physical states defined by the unitary embedding
\begin{subequations}\label{eq:transf}\begin{align}
    {\cal H}_{{\rm gr},B}\ &\rightarrow {\cal H}_{\rm gr}  &
    \psi\ &\mapsto\ B^{-\frac{1}{2}}\psi \ .
\tag{\ref{eq:transf}}\end{align}\end{subequations}
The transformation does not affect the representation used in the
sLQC case in Appendix \ref{sec:zero} because that representation is
defined directly in $\Hgr$. {In consequence $\Theta$ is still well
  defined.}

An energy density operator should be given by a suitable
symmetrization `$\vdots$' of the following operator product
\begin{equation}
  \hat{\rho}= -\vdots\frac{\rhoc}{8}
  B(v)^{\frac{1}{2}}(h_{-2}-h_{2})\tilde{A}(v)
  (h_{-2}-h_{2})B(v)^{\frac{1}{2}}\vdots\,
\end{equation}
where $\rhoc:=\sqrt{3}/(16\pi^2\gamma^2G^2\hbar)$ is the critical
energy density defined in \cite{aps-imp}.

We will start our discussion with the simplest possible case and
increase the level of complexity. To begin with let us
consider the $\Lambda=0=k$ case (see (\ref{cgr})), and
$\epsilon\not=0$. Also, as the functions $A$, $B$ we take the simplest
ones -- corresponding to sLQC
\begin{subequations}\label{eq:AB-sLQC}\begin{align}
  A\ &=\ A_{\rm sLQC}\ =\ 2A_0|v| \ ,  &
  B\ &=\ B_{\rm sLQC}\ =\ \frac{1}{|v|} \ .
\tag{\ref{eq:AB-sLQC}}\end{align}\end{subequations} Defining the
ordering ``all the functions in between the holonomy operators'', we
get the following operator
\begin{equation}\label{rhoideal}
  \hat{\rho}\ =\ -{\frac{\rhoc}{4}(h_2-h_{-2})^2} \ .
\end{equation}
The spectrum of this operator is $[0,\,\rhoc]$. Notice,
that $\rhoc$ is exactly the maximal value achieved by
$\langle\hat{\rho}\rangle(T)$ during the Big-Bounce.

Now, consider the case
\begin{subequations}\label{eq:AB-APS}\begin{align}
  A\ &=\ A_{\rm APS} \ ,   &  B\ &=\ B_{\rm APS} \ .
\tag{\ref{eq:AB-APS}}\end{align}\end{subequations}
Here we can set the same ordering as in \eqref{rhoideal},
\begin{equation}\label{eq:rho-ideal}
  \hat{\rho}_1\ =\ -{\frac{\rhoc}{8}}
  (h_{-2}-h_{2})B(v)\tilde{A}(v)(h_{-2}-h_{2}) \ ,
\end{equation}
although there is a multitude of other possibilities available, like
for example
\begin{equation}\label{eq:rho-used}
  \hat{\rho}_2\ =\ -\frac{\rhoc}{8}
  B(v)^{\frac{1}{2}}(h_{-2}-h_{2})\tilde{A}(v)
  (h_{-2}-h_{2})B(v)^{\frac{1}{2}} \ .
\end{equation}
In either case the resulting operator is \eqref{rhoideal} plus a
compact operator. Therefore the essential spectrum is
still $[0,\,\rhoc]$, however there maybe a non-empty point spectrum
above $\rhoc$ (with possible accumulation point $\rhoc$). The presence
and structure of the point spectrum depends on particular ordering
chosen. However, if the operator $\hat{\rho}$ is not explicitly
bounded from above by $\rhoc$, this opens the possibility that the
energy density expectation value can a-priori exceed $\rhoc$ for some
physically interesting states. We address this issue now.

By a calculation similar to that in \cite{sa}, one can show that
the asymptotic behaviour of the solution to equation
\begin{equation}\label{eq:eig-eq}
  \rho^{\dagger}_2\psi\ =\ \lambda\psi \ ,
\end{equation}
in dual space is the following ($\alpha,\alpha'$ are some
coefficients depending only on $\lambda$)
\begin{equation}\label{eq:limits}
  \psi(v)\ =\ \begin{cases}
    \alpha e^{i\gamma v}+\alpha'e^{-i\gamma v}+\bar{O}(v^{-1})
    \ , & \lambda<\rhoc \ , \\
    \alpha e^{\gamma v}(1+\bar{O}(v^{-1}))\ {\rm
      or}\ \alpha'e^{-\gamma v}(1+\bar{O}(v^{-1}))
    \ , & \lambda>\rhoc \ ,
  \end{cases}
\end{equation}
where $\bar{O}(v^{-n})$ is a {\it bounded} rest term decaying like
$v^{-n}$ for large $|v|$, and $\gamma$ satisfies
\begin{subequations}\begin{align}
  \label{eq:gamma1}
    \rhoc\sin^2 \gamma\ &=\ \lambda \ , \quad \lambda<\rhoc \ ,\\
  \label{eq:gamma2}
    \rhoc\sinh^2 \gamma\ &=\ \lambda \ , \quad \lambda>\rhoc \ .
\end{align}\end{subequations}
This asymptotic behavior is valid for $\lambda\not=\rhoc$ and
gives exponential decay of eigenfunctions with eigenvalues
$\lambda>\rhoc$.

It is also worth noting, that (at least for orderings given by
(\ref{eq:rho-ideal}, \ref{eq:rho-used})) there are no normalizable
solutions to \eqref{eq:eig-eq} with $\lambda=0$, therefore
$\hat{\rho}$ is invertible.

\begin{figure}[tbh!]\begin{center}
  \includegraphics[width=0.8\textwidth]{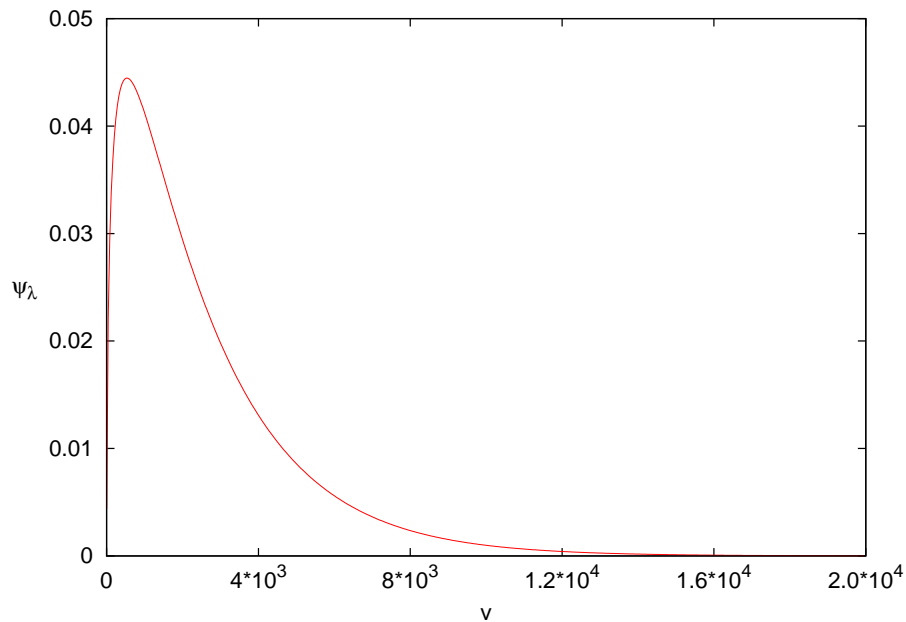}
  \caption{A normalizable eigenfunction {$\psi_\lambda$} of the
    $\hat{\rho}$ operator defined via \eqref{eq:rho-used}
    corresponding to eigenvalue $\lambda>\rhoc$. An example presented
    here has eigenvalue very close to $\rhoc$, namely
    $(\lambda-\rhoc)\approx 7.7\cdot 10^{-7}\rhoc$.}
  \label{fig:norm-eig}
\end{center}\end{figure}

At this moment we cannot exclude possibilities of existing
eigenfunctions of $\hat{\rho}$ with $\lambda>\rhoc$. Indeed, the
numerical check performed for $\hat{\rho}$ given by
\eqref{eq:rho-used} (via methods used for analysis of the spectrum of
$\Theta$ operator in \cite{apsv-spher,negL}) revealed the existence of
such eigenfunctions. An example of one of them is shown on
Fig. \ref{fig:norm-eig}. Nevertheless we showed that any
eigenfunction of $\lambda>\rhoc$ decays exponentially in $v$, with the
exponent growing logarithmically with $(\lambda-\rhoc)$ (see
\eqref{eq:gamma2}). On the other hand the numerical simulations show
that for large energies eigenfunctions of $\Theta$ are supported away
from small values of $v$. In consequence their scalar product with
eigenvalues of $\hat{\rho}$ under consideration is very small. This
fact explains why the influence of the latter cannot be observed.

An important difficulty emerges in the case of $\Lambda>0$. Then,
the scalar field energy density operator is not any longer
non-negative in $\Hil_{\rm gr}$, because it is of the form
\begin{equation}
  \hat{\rho}= -\vdots\frac{\rhoc}{8}
  B(v)^{\frac{1}{2}}(h_{-2}-h_{2})\tilde{A}(v)
  (h_{-2}-h_{2})B(v)^{\frac{1}{2}}\vdots\ -\
  \Lambda \frac{1}{V_o}\hat{V}\hat{B} \ ,
\end{equation}
{and the essential part of its spectrum is shifted with respect to
  $\Lambda=0$ case by $-8\pi G\Lambda$.}
However, the solutions of the quantum constraint (\ref{eq:c-gen})
take values in the subspace of $\Hil_{\rm gr}$ corresponding to the
non-negative part of the spectrum of $\Theta$. The question (we do
not know the answer to) is whether or not $\hat{\rho}$ restricted to
that physical subspace becomes non-negative. Since the answer
probably depends on the choice of the density operator, the
non-negativity is a condition of the ordering in a definition of
$\hat{\rho}$ consistent with the definition of $\Theta$.

\section{Concluding remarks}

The problem which we leave open is whether or not the quantum
spacetime metric tensor operator should be uniquely defined in QG.
{The proposal of such construction was made in
  Section \ref{sec:qmetric} however it suffers the factor ordering
  ambiguity.}
{At this point there are two aspects of that issue which are worth
  commenting.}

 {First,} the ordering ambiguity is restricted by the group
averaging techniques. Since the starting point for  {that procedure}
is the kinematical Hilbert space, the operator $\hpp$ commutes with
the geometry operators. Then the ambiguity is restricted {just} to a
symmetrization of the product  {$\hpp\delta(\hat{T}-T_0)$}. We
{provide an extended explanation} of that point in \cite{ga}.

 {Second,} {it is possible to} derive the
propagation equation for a quantum test field on the quantum
geometry background \cite{qft-qsp}. Not surprisingly, the result
involves the quantum metric tensor components. Remarkably, the
expression is uniquely defined, whether the quantum metric tensor
{operator itself exists} or not. Thus, the possible physical
solution of that issue may be that the quantum metric is defined uniquely
only through matter propagating on it.

Another unsolved issue is how to understand the space of solutions
to the quantum scalar constraint corresponding to the lapse function
$N=1$. {The properties of each of the individual quantum
constraint operators $\hat{C}(V)$ and, respectively, $\hat{C}(1)$
are familiar from the Schroedinger quantum mechanics. In the first
case the classical trajectories are not complete  in the evolution
parameter, namely infinite volume is achieved in finite time. That
is usually an indication that a self-adjoint extension is likely to
be not unique. In the second case, the infinite volume is achieved
only in infinite time. (The classical trajectories are incomplete at
the zero volume as well, but in LQC this does not cause any
evolution unbiguity.) Therefore there is the analogy with the
quantum mechanics. The difference, and a new ambiguity is, that in
gravity we can have two characteristics in a single theory depending
on choice of the evolution parameter. The details of the
construction of the solutions to the constraint operator
$\hat{C}(1)$ from the solutions of the variuos extensions of the
constraint operator $\hat{C}(V)$} will be presented in \cite{ga}.
Calculation of the partial observables {might}
bring even more surprises.

Finally, in this work we have considered a simplest LQC model.
However the full Quantum Gravity can be given a similar structure if
{it is formulated} according to the Brown-Kuchar model
\cite{BK,GiesThiem-algiv}. Therefore many results discussed in this
paper is likely to admit generalizations to the {full} QG.

\section*{Acknowledgments} We would like to thank Abhay Ashtekar and
Guillermo Mena-Marug\'an for extensive discussions and helpful
comments. We also profited from discussions with Martin Bojowald,
Alex Corichi, Bianca Dittrich, Marcin Domaga{\l}a, Kristina Giesel,
Carlo Rovelli, Parampreet Singh, {\L}ukasz Szulc and Thomas
Thiemann. We are grateful to the referee for his important
comments. The work was partially supported by the Polish
Minis\-terstwo Nauki i Szkolnictwa Wyzszego grant 1 P03B 075 29 and
grant 182/N-QGG/2008/0, by 2007-2010 research project N202 
n081 32/1844 , the National Science Foundation (NSF) grant
PHY-0456913 and by the Foundation for Polish Science grant
``Master''. TP acknowledges financial aid provided by the I3P
framework of CSIC and the European Social Fund, and the funds of
Polish Academy of Sciences (PAN).

\appendix

\section{Evolution operator in LQC: rigorous definition}
\label{sec:zero}

In this appendix we present the completion of the definition of the
symmetric evolution operator $\hat{\Theta}$ introduced in
(\ref{eq:Theta}), taking special care of the difficulties related with
either the vanishing
\begin{equation}
  B(0)\ =\ 0\label{B00}
\end{equation}
(in the case of $B=B_{\rm APS}$) or divergence
\begin{equation}
  B(0)\ =\ \infty\label{B0infty}
\end{equation}
(in the $B=B_{\rm sLQC}$ case).

Let us begin with (\ref{B00}). The operator $\hat{\Theta}$ has been
well defined in Sec \ref{sec:LQC} in every subspace
$\Hil_{\epsilon}\subset\Hgrb$ and ${\epsilon\in (0,4)}$ through
formula (\ref{eq:Theta}) already. The remaining case is $\epsilon=0$
(this problem is solved in \cite{aps-imp} however it is not spelled
out).

We start with a more suitable form of the constraint operator,
namely we consider the solutions to the equation
\begin{equation}
  -\frac{1}{2}\widehat{V^{-1}}\frac{\partial^2}{\partial
    T^2}\psi_T\ =\ \hat{C}_{\rm gr}\psi_T,\label{eq:constr}
\end{equation}
where $\psi_T\in {\cal H}_{\rm gr}$ and the action of
  $\Cgr$ can be written (following (\ref{cgr})) as
\begin{equation}\label{eq:Cg}
  [\hat{C}_{\rm gr}\psi_T](v)\ =\
  C^+(v)\psi_T(v+4)+C^o(v)\psi_T(v)+C^-(v)\psi_T(v-4) \ ,
\end{equation}
with $C^{o,\pm}(v)$ being real functions, of which $C^{o}(v)<0$.

Taking the scalar product of the left and the right hand sides
respectively with the vector $|0\rangle\in\Hgr$ we find
\begin{equation}
  \langle 0|\Cgr\psi_T\rangle \ =\ 0 \ .
\end{equation}
This is a condition that has to be satisfied by $\psi_T\in\Hgr$ at
every value of $T$. The meaning of this observation is, that the
functions $T\mapsto \psi_T\in\Hgr$ which satisfy the constraint
equation (\ref{eq:constr}) in fact take values only in the subspace
$\Hil_{\langle 0|\Cgr\cdot\rangle=0}$ defined by the constraint
\begin{equation}
  \langle 0|\Cgr\psi\rangle \ =\ 0 \ .
\end{equation}
However the subspace is not preserved by $\Cgr$. It (the
intersection of the domain $\Span(|v\rangle\ :\ v\in\mathbb{R})\cap
\Hil_{\langle 0|\Cgr\cdot\rangle=0}$) is mapped into another
subspace $\Hil_{\langle 0|\cdot\rangle=0}$ defined by the constraint
\begin{equation}
  \langle 0|\psi\rangle\ =\ 0 \ .
\end{equation}

On the other hand, the orthogonal projection
\begin{equation}
  \Hgr\ \rightarrow\ \Hil_{\langle 0|\cdot\rangle=0}
\end{equation}
maps isometrically
\begin{equation}
  \Hil_{\langle 0|\hat{C}_{\rm gr}\cdot\rangle=0}\ \rightarrow\
  \Hil_{\langle 0|\cdot\rangle=0} \ .
\end{equation}
This isomorphism can be used to push forward the operator $\Cgr$ to
$\Hil_{\langle 0|\cdot\rangle=0}$. An action of the resulting operator
(preserving $\Hil_{\langle 0|\cdot\rangle=0}$) is given by
\begin{equation}\label{eq:c-gen}
  [\hat{\tilde{C}}_{\rm gr}\psi](v)\ =\ \begin{cases}
    C^+(v)\psi(v+4) + C^o(v)\psi(v)
    + C^-(v)\psi(v-4) \ , &  v\not\in\{-4,0,4\} \ , \\
    \left\{ \begin{split}
      &C^+(4)\psi(8) + C^o(4)\psi(4) \\
      &- C^-(4)\left[\frac{C^+(0)}{C^o(0)}\psi(4)
      + \frac{C^-(0)}{C^o(0)}\psi(-4)\right]
    \end{split} \right\} \ , & v=4 \ , \\
    \qquad\qquad\qquad\qquad\quad 0 \ , & v=0 \ , \\
    \left\{ \begin{split}
      &C^-(-4)\psi(-8) + C^o(-4)\psi(-4) \\
      &- C^+(-4)\left[\frac{C^-(0)}{C^o(0)}\psi(-4)
      + \frac{C^+(0)}{C^o(0)}\psi(4)\right]
    \end{split} \right\} \ , & v=-4 \ , \\
  \end{cases} \ .
\end{equation}

It is worth to be stressed, that the Hilbert space isomorphism is not
unitary in the kinematical Hilbert product, but it becomes unitary,
after one endows the Hilbert space ${\cal H}_{\rm gr}$ with the
$\langle\cdot|B\cdot\rangle$ product.   Finally, the APS constraint
is imposed on functions $T\mapsto\psi_T\in {\cal H}_{\langle
0|\cdot\rangle=0}={\cal H}_{{\rm gr},B}$ and it is
\begin{equation}\label{eq:evo-gen}
  \partial^2_{T}\psi_T\ =\
  2V_oB(\hat{V})^{-1}\hat{\tilde{C}}_{\rm gr}{\psi}_T\
  =: \ -2V_o\hat{\Theta}\psi_T\ .
\end{equation}
This extends the definition of the operator $\hat{\Theta}$ given in
the previous section to the subspace  ${\cal H}_{\rm \epsilon=0}$ in
the sub-domain
$$\Span(|n\rangle\ \, : \, 0\not= n\in 4\mathbb{Z})\
\subset {\cal H}_{\rm \epsilon=0}. $$ The operator $\hat{\Theta}$ is
defined in the Hilbert space ${\cal H}_{{\rm gr},B}$ in the domain
$\Span(|v\rangle\,:\,\mathbb{R})$ (however the zero volume
vector $|0\rangle$ has zero norm in this space). It is a symmetric
operator which may have inequivalent self-adjoint extensions (see
section \ref{sec:C-noneq}), and one of them has to be chosen to make the quantum
constraint equation well defined.

Exactly that method was used in the APS papers to study the physical
solutions which take values in the subspace ${\cal H}_{\epsilon=0}$
(solutions preserved by the reflection $P|v\rangle=|-v\rangle$).
\medskip

Let us turn now to the (\ref{B0infty}) case. Now the problem is in
introducing the scalar product $\langle\cdot|B\cdot\rangle$. However
instead, we can modify the procedure of going from (\ref{eq:constr})
to an analog of (\ref{eq:evo-gen}) defining the operator
\begin{equation}
  \hat{\Theta}\ :=\ - B(\hat{V})^{-1/2} \Cgr B(\hat{V})^{-1/2} \ ,
\end{equation}
which is well defined and symmetric (in the domain ${\rm
  Span}(|v\rangle\ :\ v\in\mathbb{R})$) with respect to the original
kinematical inner product of $\Hgr$.

\end{document}